\documentclass[pra, aps,letterpaper, twocolumn,  preprintnumbers,superscriptaddress]{revtex4}
\usepackage{enumerate}
\usepackage{epsfig}
\usepackage{verbatim}
\usepackage{amsmath}
\usepackage{latexsym} 
\usepackage{revsymb} 

\usepackage{ifthen} 
\usepackage{mathrsfs}

\usepackage{natbib}
\usepackage{amsfonts}
\usepackage{amsmath}
\usepackage{amssymb}

\usepackage{amsthm}
\usepackage{graphicx}

\newcommand*{\textfrac}[2]{{{#1}/{#2}}}

\newcommand*{\cC}{\mathcal{C}}
\newcommand*{\cD}{\mathcal{D}}

\newcommand*{\cH}{\mathcal{H}}

\newcommand*{\cE}{\mathscr{E}}

\newcommand*{\tr}{\mathop{\mathrm{tr}}\nolimits}

\newcommand{\poly}{\mathsf{poly}}

\newcommand{\ket}[1]{|#1\rangle}

\newcommand{\proj}[1]{|#1\rangle\langle#1|}

\newcommand*{\myproj}[1]{P_{#1}}

\newcommand*{\myapprox}[1]{\ \widetilde{\mapsto}_{#1}\ }
\newcommand*{\arrangementfirst}{\raisebox{-50.886pt}{\epsfig{file=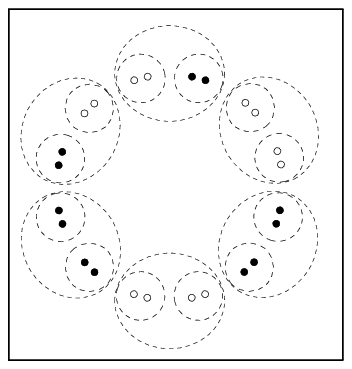,clip=}}}

\newcommand*{\injectionweave}{\raisebox{-25.383pt}{\epsfig{file=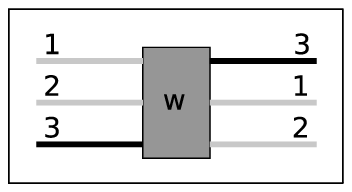,clip=}}}

\newcommand*{\injectionweaverecursive}{\raisebox{-31.086pt}{\epsfig{file=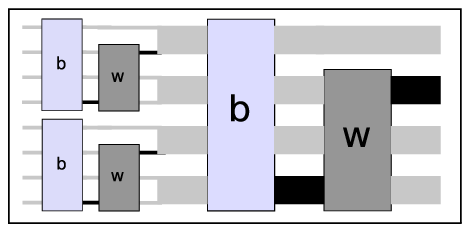,clip=}}}

\newcommand*{\arrangementsecond}{\raisebox{-50.886pt}{\epsfig{file=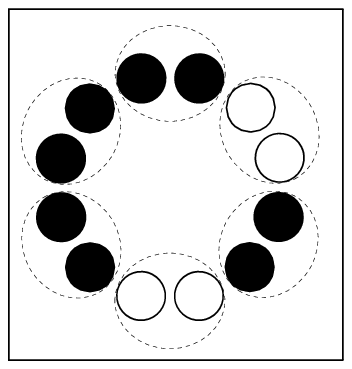,clip=}}}

\newcommand*{\boxbasisfirst}{\raisebox{-14.043pt}{\epsfig{file=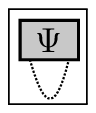,clip=}}}

\newcommand*{\boxbasissecond}{\raisebox{-14.043pt}{\epsfig{file=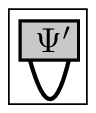,clip=}}}

\newcommand*{\boxbasisthird}{\raisebox{-14.043pt}{\epsfig{file=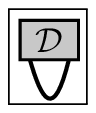,clip=}}}

\newcommand*{\bubble}{\raisebox{-11.943pt}{\epsfig{file=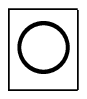,clip=}}}

\newcommand*{\Rmatrixa}{\raisebox{-10.503pt}{\epsfig{file=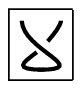,clip=}}}

\newcommand*{\Rmatrixaprime}{\raisebox{-10.503pt}{\epsfig{file=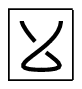,clip=}}}

\newcommand*{\Rmatrixb}{\raisebox{-16.203pt}{\epsfig{file=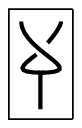,clip=}}}

\newcommand*{\Rmatrixbprime}{\raisebox{-16.203pt}{\epsfig{file=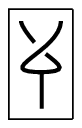,clip=}}}

\newcommand*{\Rmatrixresolvedb}{\raisebox{-16.203pt}{\epsfig{file=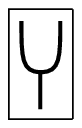,clip=}}}

\newcommand*{\Rmatrixresolveda}{\raisebox{-10.503pt}{\epsfig{file=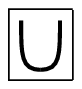,clip=}}}

\newcommand*{\Dssdd}{\raisebox{-14.043pt}{\epsfig{file=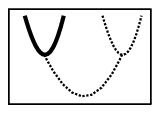,clip=}}}

\newcommand*{\msigma}{\raisebox{-19.026pt}{\epsfig{file=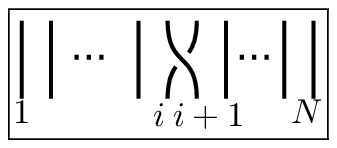,clip=}}}

\newcommand*{\msigmainv}{\raisebox{-19.026pt}{\epsfig{file=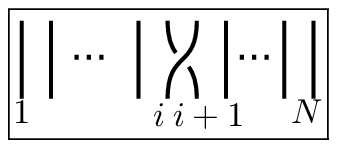,clip=}}}

\newcommand*{\Dddss}{\raisebox{-14.043pt}{\epsfig{file=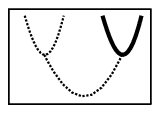,clip=}}}

\newcommand*{\Fmoveaone}{\raisebox{-18.306pt}{\epsfig{file=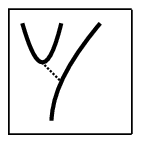,clip=}}}

\newcommand*{\Fmoveatwo}{\raisebox{-18.306pt}{\epsfig{file=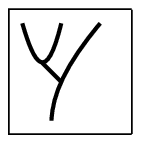,clip=}}}

\newcommand*{\Fmovebone}{\raisebox{-18.306pt}{\epsfig{file=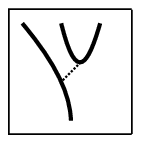,clip=}}}

\newcommand*{\Fmovebtwo}{\raisebox{-18.306pt}{\epsfig{file=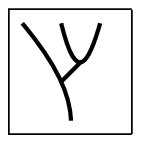,clip=}}}

\newcommand*{\Dssss}{\raisebox{-14.043pt}{\epsfig{file=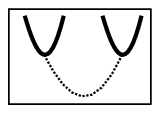,clip=}}}

\newcommand*{\Sssss}{\raisebox{-14.043pt}{\epsfig{file=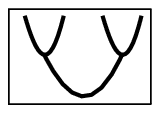,clip=}}}

\newcommand*{\Ssdds}{\raisebox{-14.043pt}{\epsfig{file=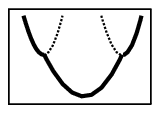,clip=}}}

\newcommand*{\Sdssd}{\raisebox{-14.043pt}{\epsfig{file=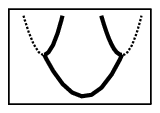,clip=}}}

\newcommand*{\basisPsi}{\raisebox{-14.043pt}{\epsfig{file=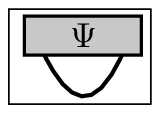,clip=}}}

\newcommand*{\basisPsisec}{\raisebox{-14.043pt}{\epsfig{file=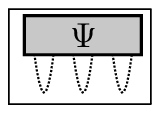,clip=}}}

\newcommand*{\basisPsithird}{\raisebox{-14.043pt}{\epsfig{file=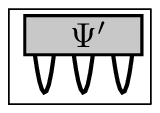,clip=}}}

\newcommand*{\basisPsifourth}{\raisebox{-14.043pt}{\epsfig{file=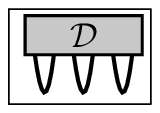,clip=}}}

\begin{document}

\title{Composite anyon coding and the initialization of a topological quantum computer}
\author{Robert \surname{K{\"o}nig}}
\email[]{rkoenig@caltech.edu}
\affiliation{Institute for Quantum Information, Caltech, Pasadena CA 91125, USA}
\begin{abstract}
Schemes for topological quantum computation are usually based on the assumption that the system is initially  prepared in a specific state. In practice, this state preparation is expected to be challenging  as  it  involves non-topological operations which heavily depend on the experimental realization and are not naturally robust against noise. Here we show that this assumption can be relaxed by using composite anyons: starting from an unknown state with reasonable physical properties, it is possible to efficiently distill
suitable initial states for computation purely by braiding, i.e., reversible gates. This is achieved by encoding logical information in a subsystem code with  gauge system corresponding to the internal degrees of freedom of composite objects. 
\end{abstract}
\maketitle
\section{Introduction}
The  effects of imperfections and decoherence are arguably the most serious obstacle faced when trying to design a quantum computer. Their study has led to the development of elaborate schemes for fault-tolerant quantum computation. These efforts have culminated in the {\em threshold theorem}~\cite{AhBO97,KLZ98,Preskill98error}, which  establishes that reliable computation is in principle possible given a sufficiently low error probability of each individual gate. A central idea underlying these results is the concatenation of codes, first formalized by Forney~\cite{For66} and later generalized to quantum error-correcting codes~\cite{Got96,Got97}. Here, logical information is encoded in a recursive fashion, where the qubits constituting a codeword are themselves represented by an error-correcting code. This leads to an exponential reduction of the effective error rate in the number of concatenation levels.

\vspace{-0.035cm}

{\em Topological quantum computation} (TQC) is an alternative (and in some sense complementary) approach to fault-tolerance~\cite{Kitaev03,preskill,Freedmanetal03,Nayaketal08}.  The basic steps for TQC can roughly be summarized as follows (but see~\cite{bonfreenay08} for an alternative complete set of basic operations): (i)~create a certain initial state of non-abelian anyons (e.g., the state of many particle-antiparticle pairs, as explained below), (ii)~braid the anyons around each other to perform gates, and (iii)~measure the resulting state (that is, the total topological charge of a region). The non-local nature of the encoding used here is expected to lead to significantly lower error rates compared to more standard implementations. Moreover, 
gates are intrinsically robust due to the fact that~(ii) only involves topological operations.
\vspace{-0.035cm}

Unfortunately, steps~(i) and (iii) of TQC are generally system-dependent, and require performing non-topological operations which are not naturally protected against noise. For example, measurement of topological charge involves making interferences from the observation of a non-topological quantity (e.g., current) through approximate relations valid only in the given experimental setup.
Here we focus on the state preparation problem~(i). We give a scheme for creating a suitable initial state purely by braiding, starting from an unknown state supported on a physically relevant subspace. This reduces TQC to a sequence of braids followed by measurement. It  eliminates  all unprotected operations apart from those at the end of the computation. 

The basis for our scheme is the fact that a region of a topologically ordered 2D-quantum medium can be regarded as a dot where a composite anyon resides. This composite object carries the total (topological) charge of all the anyons contained in that region. Anyon systems therefore provide a natural mechanism for concatenated coding: fusion basis states of composite objects can be used to encode information. At each level of concatenation, a number of subregions are joined into a region forming a new composite object. Ignoring the degrees of freedom within subregions, the resulting encoding has the structure of a subsystem code~\cite{KnillLafl97,KniRayVio00,Bacon06}, where the Hilbert space decomposes as~$\cH=(\cC\otimes\cD)\oplus\cE$, and the logical information is encoded in~$\cC$. Here~$\cD$ describes the internal degrees of freedom of the composite anyons. Importantly, computations in the code space~$\cC$ can be performed in the same manner as for the bare anyons. In particular, the complexity of braiding two regions around each other only scales polynomially with the size of their boundary. (This can be seen explicitly for schemes based on local code deformations.) A similar statement can be made for the  measurement of composite topological charge, which can be achieved by anyon interferometry~\cite{Bonetal08}, for example. To compute, it is therefore sufficient to be able to prepare any state supported on the subspace~$\ket{\Psi_{init}}\otimes\cD$, where $\ket{\Psi_{init}}$ is a state of a fixed number of anyons representing the logical state~$\ket{0}^{\otimes n}$ of~$n$~computational qubits. 

We consider (mixed) states supported on a physically relevant subspace. We show that  any such state~$\rho$  can be turned into a state supported on $\ket{\Psi_{init}}\otimes\mathcal{D}$ using reversible gates. Moreover, we construct braid sequences implementing this efficiently.  In other words, we construct a distillation procedure which concentrates the entropy of~$\rho$ into~$\cD$. Using this distillation procedure circumvents the need for measurements when preparing the initial state for computation. 

The remainder of this paper is structured as follows: In Section~\ref{sec:anyonbackground}, we give some background on anyons and explain the action of the braid group.  In Section~\ref{sec:compositeanyondistillation}, we present our scheme for distillation of composite anyons. We conclude in Section~\ref{sec:conclusions}.

\section{Anyons and the braid group action\label{sec:anyonbackground}}
\subsection{FQHE and qudit lattice Hamiltonians}
The fractional quantum Hall (FQH)  effect  is the most well-known example of a topologically ordered state of matter. At filling fraction $\nu=5/2$, this system is believed to be in the universality class of the Moore-Read Pfaffian state~\cite{MooreRead}, which is the $k=2$ state in the Read-Rezayi-sequence of states~\cite{readrezayia,readrezayib}. Similarly, for $\nu=12/5$, the system is believed to be described by the~$k=3$ Read-Rezayi-state. Quasi-particle excitations above these states are superconducting vortices carrying fractional (electric) charge~$e/4$. They are predicted theoretically to correspond to the anyons described by $\mathfrak{su}(2)_k$~Chern-Simons-Witten theory. In  principle, these realize two of the most basic non-abelian phases: the Ising TQFT (for $k=2$) and the Fibonacci TQFT (for $k=3$).  At present, though, there is only some experimental evidence   for the existence of these quasiparticles, and it is restricted to the former case~\cite{dolevetal,raduetal,willettetal}.

FQH systems are not the only candidate systems for building a topological quantum computer. Of particular interest are qudit lattice Hamiltonians with topological order. Examples are Kitaev's toric code~\cite{Kitaev03}, Kitaev's honeycomb model~\cite{KitaevAnyons} or Levin- and Wen's  Hamiltonian~\cite{LevinWen}. While these Hamiltonians are unlikely to occur in nature, it is in principle possible to engineer systems with corresponding interactions, and some experimental proposals for doing so exist (see e.g.,~\cite{duanetal03,michelietal06}). With an appropriate choice of Hamiltonian, anyons corresponding to any ``doubled'' (non-chiral) TQFT can be obtained from such a construction. This approach to achieving topological order also has an additional advantage:  the  local Hamiltonian  terms can be interpreted as local stabilizers of a quantum error-correcting code of qudits. Using this code for storage and processing of quantum computation merely requires the ability to measure syndromes locally, instead of requiring a system described by the Hamiltonian. In this setting, the problem of decoherence at non-zero temperature (see e.g.,~\cite{Alickietal09}) can be overcome by performing active error-correction~\cite{Dennisetal02}. Indeed, this gives rise to excellent threshold results for quantum error-correction~\cite{Raussendorfetal05,Raussendorfharrington06}.

To use a topologically ordered medium for computation, one requires the experimental ability to manipulate, i.e., create, move (braid) and measure anyons. For FQH states, proposals for implementing anyonic charge measurements are most developed.  These employ  beams of probe quasiparticles (e.g., constituting a current along an edge), which are  sent through an interferometric setup (e.g., by deforming the edge around a region to be measured).  Because of the non-abelian statistics,  topological charge information can be deduced from the interference pattern observed by detecting whether a quasiparticle is present at the outputs of the interferometer. Since the latter detection requires coupling to e.g., the electric charge of the probe anyons, this is a rather indirect procedure which is susceptible to noise. Nevertheless, this is at present the most promising experimental approach to observing anyonic statistics. In contrast to measurements,  current theoretical proposals in the FQH setting for other operations, such as controlled movement of individual particles, perhaps have little chance of actually being realized.

The manipulations required to perform a topological quantum computation in a qudit lattice system consist of local unitaries and measurements acting on a small number of qudits at a time.  This has been studied in detail for the toric code~\cite{Kitaev03} and the Levin-Wen-Hamiltonians~\cite{LevinWen}. Anyons are created and moved by applying certain ``ribbon operators'' (which can be decomposed into locally acting operators). Some proposed schemes~\cite{Freedman00,KoeKupRei10} directly use a topologically degenerate ground space and proceed by adiabatically deforming the Hamiltonian in a local fashion.  Measurements of topological charge require decoding the encoded information, and thus are automatically subject to noise. Corresponding circuits are discussed e.g., in~\cite{Dennisetal02,KoeReiVid09,DuclosPoulin10}.

\subsection{Fibonacci anyons}
We formulate our scheme for the Fibonacci theory,  the simplest computationally universal anyon model. We proceed with a short description of the corresponding braid group representation.  For more details, we refer to the literature. Excellent introductions can be found e.g., in~\cite{preskill,Trebstetal08}.

  The Fibonacci model has only one non-trivial particle~$\tau$ with fusion rule $\tau\times\tau=1+\tau$. For simplicity, we consider a (infinite) plane with $N$~localized dots, each of which can support a $\tau$ particle (but may also have trivial topological charge). We assume that the total topological charge is~$1$. The Hilbert space then decomposes into a direct sum 
$\cH\cong \bigoplus_{n=0}^N \bigoplus_{i=1}^{\binom{N}{n}} V^{i}_n$, where $V^i_n$ is the space of $n$ $\tau$-anyons localized at dots specified by the subset~$i$. Universal quantum computation can be performed in a subspace $\mathbb{C}^{N/4}\subset V^1_N\cong:V_N$, but here we take a more general viewpoint.

The space~$V^i_N$ of $N$~$\tau$-anyons is characterized by the action of the braid group. In the FQH setting, it is the degenerate eigenspace of states with localized excitations at specified locations (i.e., the dots). In topological computation schemes based on the degenerate ground space of a qudit lattice model, the space~$V^i_N$ is a certain subspace of the ground space. Here dots correspond to defects, i.e.,  holes in the latttice structure or (possibly extended)  areas where the local Hamiltonian terms are  different from the
bulk. The mapping  between states of the physical system  and the abstract anyonic space~$V_N$ is discussed  extensively in the literature (see e.g.,~\cite{readrezayia,preskill} and~\cite{Kitaev03,KitaevAnyons,KoeKupRei10} for lattice systems). Similarly, braids correspond to certain (physical) operations on the system, as discussed above. Here we will restrict our discussion to the abstract anyonic formalism which provides a system-independent description of the action of the braid group.

We  use the standard diagrammatic representation of states and operators on $\cH$ (see e.g.,~\cite{bondersonthesis} for more details): To every tree with trivalent vertices and~$N$~leaves, we can associate an orthonormal basis of $\cH$ with elements given by different labelings of the edges by $\{1,\tau\}$ consistent with the fusion rules (i.e., with the number of edges labeled~$\tau$ incident to any vertex not equal to~$1$). We fix the label at the root of the tree to~$1$ corresponding to the trivial topological charge at the $\infty$-boundary. The labels attached to the edges connected to the leaves determine which of the spaces $V^i_n$ a labeling corresponds to. The tree specifies the order of fusing, and basis elements corresponding to different trees are related by so-called  $F$-moves. In the following diagrams, we represent edges with the label~$\tau$ by solid lines, whereas edges with the trivial label~$1$ will be dotted or omitted altogether. 
An $F$-move then is a local substitution of subgraph by a linear combination as specified by the  identities
\begin{align}
\begin{matrix}
\Fmoveaone &=\frac{1}{\phi}\Fmovebone +&\frac{1}{\sqrt{\phi}}\Fmovebtwo\\
\Fmoveatwo &=\frac{1}{\sqrt{\phi}}\Fmovebone -&\frac{1}{\phi}\Fmovebtwo
\end{matrix}\label{eq:fmoveequation}
\end{align}
where  $\phi=\frac{\sqrt{5}+1}{2}$ is the golden ratio.

The braid group $B_N$ on $N$~strands is generated by the elementary braids~$\sigma_1,\ldots,\sigma_{N-1}$ satisfying the Artin relations $\sigma_i\sigma_j=\sigma_j\sigma_i$ for $|i-j|\geq 2$ and $\sigma_i\sigma_{i+1}\sigma_i=\sigma_{i+1}\sigma_{i}\sigma_{i+1}$ for $1\leq i\leq n-2$. To describe the action of~$B_N$ on~$\cH$, one represents the generators as
\begin{align*}
\sigma_i&=\msigma\\
\sigma_{i}^{-1}&=\msigmainv\ .
\end{align*}
Group multiplication then corresponds to stacking such diagrams. To apply a braid~$b\in B_N$ to a basis element~$\ket{\Psi}\in\cH$, one attaches the graphical representation of $b$ to the  leaves of the tree specifying~$\ket{\Psi}$, and then reexpresses the result in the original basis using a set of local rules. These allow to arbitrarily add and remove edges with the trivial label~$1$ in addition to isotopic deformations that keep the leaves fixed, $F$-moves as in~\eqref{eq:fmoveequation}, substitutions of bubbles by scalars as specified by
\begin{align}
\bubble &=\phi \label{eq:bubblerule}
\end{align}
 and resolution of crossings using the $R$-matrix. The latter take the form
\begin{align*}
\begin{matrix}
\Rmatrixa &= e^{-4\pi i/5}\Rmatrixresolveda\qquad &\Rmatrixaprime &= e^{4\pi i/5}\Rmatrixresolveda\\
\Rmatrixb &=e^{3\pi i/5} \Rmatrixresolvedb\qquad &\Rmatrixbprime &=e^{-3\pi i/5}\Rmatrixresolvedb
\end{matrix}\ .
\end{align*}

The space $V^{i}_n$ 
has dimension  $\dim V^{i}_n=\frac{1}{\sqrt{5}}\left(\phi^{n-1}-(-\frac{1}{\phi})^{n-1}\right)$. It is invariant  under the action the purebraid group~$PB_N$ on $N$~strands. This is the subgroup of the braid group~$B_N$ where each strand begins and ends in the same position.  The restriction of this group action to~$V^i_n$ has a dense image in the  unitary group~$PU(\dim V^i_n)$
modulo global phase factors~\cite{FreedLarsWang02} (i.e., is computationally universal). More precisely,
$PU(M)=SU(M)/\mathbb{Z}_M$, where $\mathbb{Z}_M$ is the subgroup of $SU(M)$ generated by $e^{2\pi i/M}\cdot \mathsf{id}$.

\section{Distilling composite anyons\label{sec:compositeanyondistillation}}
\subsection{The setting}
\begin{figure}
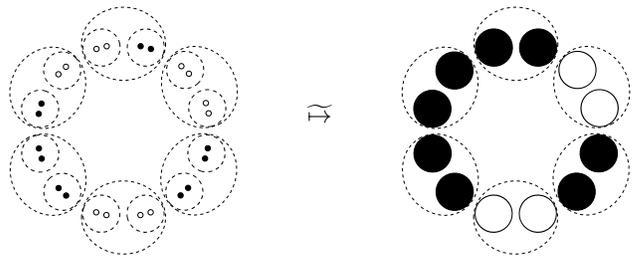

\arrangementfirst $\qquad\myapprox{}\qquad$ \arrangementsecond
\caption{Left: A possible arrangement of the dots (small circles) in the plane; we write basis states arranged from left to right, but the actual dots need not be arranged on a line. A typical state in $\cH_{phys}$ has some $\tau\tau$-pairs (depicted by filled circles), and trivial composite topological charge for any larger region containing a pair of dots (dotted lines).
Right: The result of applying the described braiding transformation $(w\circ b)^{\otimes 6}$ to each group of four is a state with composite anyon pairs in the indicated locations.\label{fig:anyonarrangement}}
\end{figure}

Let us identify physically relevant states. We envision the dots to be arranged in spatially separated pairs, see Fig.~\ref{fig:anyonarrangement}. 
Let $\ket{vac}\in V^1_0$ be the (up to a phase) unique state of no particles. We could imagine the system to be initially prepared in this vacuum state, but subsequently undergo processes which create particle-antiparticle pairs locally at pairs of dots. We  expect tunneling of (topological) charges between different pairs to be exponentially suppressed in the inter-pair distance (cf.~\cite{bonderson09}). Due to the geometric arrangement of the dots, the resulting state will therefore essentially be a mixture of superpositions  with trivial charge for each pair of dots, but having $\tau\tau$-pairs within some of them.

It is natural to ask for the distribution of $\tau\tau$-pairs in such a state. Here we will make the minimal assumption that within region~$\Omega$ containing $m_\Omega$~pairs of dots, there is a non-zero number of $\tau\tau$-pairs except with a probability exponentially small in~$m_\Omega$. More precisely, let~$\myproj{\Omega}$ 
denote the projection onto the subspace spanned by states where all pairs in $\Omega$ have trivial total topological charge and at least one $\tau\tau$-pair is present. We assume that the states are supported on the image of $\myproj{\Omega}$ up to a fraction exponentially small in~$m_\Omega$.  States of this form   are the starting point for our braiding procedure.  

As a prototypical example of a state satisfying our assumptions, suppose $\rho=\cE^{\otimes N/2}_p(\proj{vac})$ is the result of a process $\cE_p$ which creates particle pairs with probability~$p$ locally. Then the number of pairs in a region~$\Omega$ is binomially distributed with parameters $(m_\Omega,p)$, and $\tr(\myproj{\Omega}\rho)=1-(1-p)^{m_\Omega}$.

 We will assume that after the initial pair creation processes, the dots are moved apart from each other so that further particle tunneling is suppressed. 
We will show how  states satisfying our assumptions can be efficiently mapped to a subspace $\ket{\Psi_{init}}\otimes\cD$ (up to a negligible error)  by braiding, where $\ket{\Psi_{init}}\in V_{2k}$ is the state of $k$~composite $\tau\tau$-pairs each with trivial topological charge. The latter  can be identified with the state~$\ket{0}^{\otimes k/2}$ of~$k/2$~logical qubits in a standard encoding.

We conclude this section by commenting on the physical systems where our scheme may be applicable. Because  of the restricted repertoire of currently realizable operations in the FQHE setting, the proposed procedure may be more useful in the lattice Hamiltonian setting (if such a system can be realized). Furthermore, in quantum Hall states, anyons typically carry additional quantum numbers such as electric charge. 
 This should simplify their detection by local measurements; it also leads to an additional energetic cost for having anyon pairs. In contrast, anyons encoded in the ground space of a lattice system (as in~\cite{Freedman00,KoeKupRei10}) have degenerate energies even  for different topological charges. Also, these anyons are not naturally localized, but correspond to topologically non-trivial loops around holes. To measure these topological charges requires non-local measurements which are susceptible to noise. Thus such lattice systems are natural candidates for the application of the proposed scheme.

\subsection{Distillation of a single pair}
For simplicity, we first consider the problem of ``distilling'' a single composite $\tau\tau$-pair, i.e., $k=1$, and generalize to $k>1$ later. We show the following: Consider a region $\Omega=\Omega^A\cup\Omega^B$ partitioned into disjoint sets $\Omega^A,\Omega^B$ each containing $m_\Omega/2=m_{\Omega^A}=m_{\Omega^B}$ pairs of dots. Then there is an efficiently computable sequence of $\poly(m_\Omega,\log\textfrac{1}{\varepsilon})$ elementary braiding operations within~$\Omega$ with the following effect: the image $\ket{\Psi'}$ of any state $\ket{\Psi}\in \myproj{\Omega}\cH$  can be approximated in trace norm to distance~$\varepsilon$ by a superposition of states with a composite $\tau\tau$-pair between $\Omega^A$ and $\Omega^B$.  Pictorially,
\begin{align}
\boxbasisfirst \myapprox{\varepsilon}\boxbasissecond\in\boxbasisthird
\label{eq:usefulmone}
\end{align}
where the notation $\myapprox{\varepsilon}$ indicates that the result is approximate up to an error~$\varepsilon$ in trace norm. This picture is to be understood in a basis where $\Omega^A$ and $\Omega^B$ are the leaves of two subtrees joined together. In other words, the resulting state is supported on the subspace $\ket{\Psi_{init}}\otimes\cD\subset \cC\otimes\cD$, where  the code space $\cC\cong V_{2}\cong\mathbb{C}$ is the space of $2$~composite $\tau$-anyons defined by the subregions $\Omega^A,\Omega^B$, and the space~$\cD$ corresponds  to the internal degrees of freedom in the subregions (represented by boxes). 
 Elementary braiding operations are  defined as clockwise or counterclockwise exchanges of  regions contained within $\Omega$, and thus have a polynomial complexity  in the length of the boundary of~$\Omega$. 

\begin{figure}
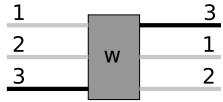

\begin{center}
\injectionweave
\end{center}
\caption{The form of an injection weave. The warp strand is shown in bold and the
box stands for a sequence of braids whose details are irrelevant here (explicit constructions of such weaves are given in~\cite{Bonesteeletal05,Hormozietal06}). The injection weave acts as the identity on the space of three $\tau$-anyons up to a small error. \label{fig:injectionweave}}
\end{figure}

To establish our main claim, consider the special case where $m_\Omega=2$, i.e., $\Omega^A$ and $\Omega^B$ each contain a pair of dots. Here the subspace $\myproj{\Omega}\cH$  is spanned by states which locally have the form of the three states 
 \begin{align*}
\Dssdd,\Dddss,\Dssss\ .
\end{align*}
Our goal is to map such states to states of the form
\begin{align}
\basisPsi\label{eq:solidPsi}
\end{align}
by an appropriate braid of the dots. By the density result mentioned above (applied to the space $V_4$), there is an element~$b\in PB_4$ which approximately maps
\begin{align}
\Dssss \mapsto\myapprox{\varepsilon} \Sssss\  \label{eq:approxmapv}
\end{align}
up to an irrelevant phase. Using the fact that $b$~is a purebraid, we conclude that the other two states are mapped to themselves up to phase factors, i.e.,
\begin{align}
\Dddss\mapsto e^{i\varphi_1}\Dddss, \ \Dssdd \mapsto e^{i\varphi_2}\Dssdd\ \label{eq:invariancemapping}
 \end{align}
 Note that, by the Kitaev-Solovay theorem~\cite{KitaevShenbook}, a sequence of $C(\log \textfrac{1}{\varepsilon})^\alpha$ (where $C$ is a constant and $\alpha\approx 4$) generators implementing a braid~$b$ with the desired property can be found efficiently. Executing this sequence is the first step in our procedure.
It remains to map the two states in~\eqref{eq:invariancemapping} to states of the form~\eqref{eq:solidPsi}, {\em while leaving the rhs.~of~\eqref{eq:approxmapv} invariant}. For this purpose, we use the {\em injection weave} construction proposed in~\cite{Bonesteeletal05}. This is an element $w\in B_3$ with the following properties: (i)~up to an error~$\varepsilon$, it acts as the identity on the space of three $\tau$-anyons,  (ii)~it is a {\em weave}, i.e., only a single strand, the {\em warp}-strand is moved around~$2$ stationary strands and (iii)~the warp strand starts in the right-most position, and is transferred to the leftmost position. Fig.~\ref{fig:injectionweave} shows the form, i.e., conditions (ii) and (iii) defining such a braid. As for the braid~$b$, an injection weave~$w$ of length~$C(\log\textfrac{1}{\varepsilon})^\alpha$ can be found  efficiently using the Kitaev-Solovay algorithm~(see~\cite{Simonetal06} for a general discussion of the existence of injection weaves and its application to the construction of ``pureweaves'' for universal computation). It is clear that applying such an element~$w$ to the three rightmost strands maps 
\begin{align*}
\Dssdd \mapsto e^{i\phi_1}\Ssdds\ ,\qquad \Dddss \mapsto e^{i\phi_2}\Sdssd\ ,
\end{align*} 
for some phases $\phi_1,\phi_2$,
while approximately preserving the state on the rhs.~of~\eqref{eq:approxmapv} up to an error~$\varepsilon$ by property~(i). Importantly, the precision~$\varepsilon$ of $b$ and $w$ can be improved exponentially by a polynomial increase in the length of the braids.
We conclude that the composition $w\circ b$ achieves the desired result in the case where $m_\Omega=2$.

\subsection{Distillation of multiple pairs}
To prove our claim in the general case, consider a region~$\Omega$ containing $m_\Omega=2^{\ell}$~pairs of dots. 
We divide these into groups of $4$~dots (two pairs), and apply $w\circ b$ to each of these groups. Picking an area around each pair, we can think of the result in terms of ``superdots'' (each containing a pair) where composite $\tau$-anyons may reside. Inspecting the previous analysis, we conclude that by going from dots to composite objects, we half the number of objects, while effectively increasing the density of $\tau\tau$-pairs. More precisely, let $\myproj{\Omega}^{\ell-1}$ be defined  in the same way as~$\myproj{\Omega}$, but for the $2^{\ell-1}$~pairs of superdots. Then the map $(w\circ b)^{\otimes 2^{\ell-1}}$ takes
\begin{align}
\begin{matrix}
\ket{\Psi}\in &\myproj{\Omega}\cH &\myapprox{ 2\varepsilon\cdot 2^{\ell-1}} &\ket{\Psi'}\in\myproj{\Omega}^{\ell-1}\cH\ ,
\end{matrix}\label{eq:inductidentityv}
\end{align}
and is composed of $2C(\log\textfrac{1}{\varepsilon})^\alpha\cdot 2^{\ell-1}$ elementary braiding operations. Here we used the triangle inequality to give an upper bound on the error.

\begin{figure}
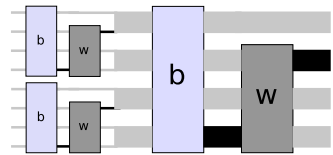

\injectionweaverecursive
\caption{The first two levels of the composite braid used for the preparation of the initial state. Thick strands correspond to composite anyons.\label{fig:circuit}}
\end{figure}

Finally, we use this procedure recursively: at the $r$-th level (with $r$ decreasing from $\ell$ to $1$), we
divide the given $2^{r+1}$~regions  into groups of four, apply $(w\circ b)^{\otimes 2^{r-1}}$ and obtain  $2^{r}$~superregions containing pairs of regions. 
This is shown schematically in Fig.~\ref{fig:circuit} (cf.~Fig.~\ref{fig:anyonarrangement}). We are interested in composite $\tau\tau$-pairs among pairs of these superregions (at most~$2^{r-1}$ may be present), and  define~$\myproj{\Omega}^{r-1}$ accordingly. 
By induction using~\eqref{eq:inductidentityv}, we conclude that  after $\ell-1$ steps, we obtain a map
\begin{align*}
\ket{\Psi}\in &\myproj{\Omega}\cH \myapprox{\delta} \ket{\Psi''}\in \myproj{\Omega}^0\cH\qquad\textrm{ with precision }\\
\delta &=(2\varepsilon)\sum_{r=1}^{\ell} 2^{r-1}<2^\ell(2\varepsilon) \ .
\end{align*} 
Furthermore, this requires $2C(\log\textfrac{1}{\varepsilon})^\alpha\cdot \sum_{r=1}^\ell 2^{r-1}<2^{\ell}\cdot 2 C(\log\textfrac{1}{\varepsilon})^\alpha$ elementary braiding operations. We conclude that any desired precision $\bar{\varepsilon}$ can be obtained with $m_\Omega\cdot \poly (\log m_\Omega,\log\textfrac{1}{\bar{\varepsilon}})$ elementary braiding operations.  Furthermore, by definition of~$P^0_\Omega$, the resulting state has a (composite) $\tau\tau$-anyon between two regions of~$\Omega$ having~$2^{\ell}$~points each. We have thus established our main claim for $k=1$ encoded $\tau\tau$-pair.

Let us now generalize this result to the generation of $k>1$ composite $\tau\tau$-pairs. Since the transformation~\eqref{eq:usefulmone} does not affect the region outside $\Omega$, we can simply repeat it $k$~times.
That is, we fix a partition $\bigcup_{\alpha=1}^k \Omega_\alpha$ of the system into sufficiently large regions~$\Omega_\alpha$, each containing~$m$ pairs of dots~($m$~is a power of $2$). As argued above, the space  $\cH_{phys}:=(\prod_{\alpha=1}^k \myproj{\Omega_\alpha})\cH$, which is the space of states having at least one $\tau\tau$-pair in each region, essentially supports all states of interest. Applying the transformation~\eqref{eq:usefulmone}  for each $\alpha$  maps any such state into a state of $k$~encoded $\tau\tau$-pairs with error at most $k\varepsilon$, i.e., (for~$k=3$)
\begin{align*}
\basisPsisec \myapprox{k\varepsilon} \basisPsithird \in  \basisPsifourth
\end{align*}  The code space~$\cC\cong V_{2k}$ is the space of $2k$~composite $\tau$-anyons defined by subregions $\Omega^A_1,\Omega^B_1,\ldots,\Omega^A_k,\Omega^B_k$, where we have chosen a bipartition $\Omega_\alpha=\Omega^A_\alpha\cup\Omega^B_\alpha$ for each $\alpha=1,\ldots,k$ with $m_{\Omega^A_\alpha}=m_{\Omega^B_\alpha}=m/2$. The space~$\cD$ represents the internal degrees of freedom in the subregions.  Note that the resulting state can be used to represent the state $\ket{0}^{\otimes k/2}$ of $k/2$~logical qubits in a ``computational'' subspace $\mathbb{C}^{k/2}\subset V_{2k}$ using a standard encoding~\cite{Freedmanetal03}.

 In summary, our braiding sequence is efficiently computable and creates $k$~composite $\tau\tau$-pairs from any state supported on~$\cH_{phys}=(\prod_{\alpha=1}^k \myproj{\Omega_\alpha})\cH$. Any desired precision~$\varepsilon$ is obtained using a polynomial number~$\poly(m,k,\log\textfrac{1}{\varepsilon})$ of elementary braiding operations.  The latter are defined as clockwise or counterclockwise exchanges of  regions contained within the sets~$\Omega_\alpha$, and thus have a polynomial complexity  in the length of the boundary of~$\Omega_\alpha$.

\section{Conclusions\label{sec:conclusions}}
We have demonstrated that using composite anyons in conjunction with specific braiding sequences can eliminate the need for preparing a particular initial state for topological quantum computation.
Our scheme is highly parallelizable, and bears a remarkable similarity with entanglement renormalization methods~\cite{ER}, suggesting that these concepts might extend to the setting of anyons.  Our work is merely a proof of principle and is restricted to Fibonacci anyons, though our arguments can be adapted to give procedures for the anyons of $su(2)_3$~Chern-Simons-Witten theory along the lines of~\cite{Hormozietal06},  as well as to e.g., the doubled Fibonacci theory. Whether or not similar schemes can be constructed for more general anyon models is an interesting open problem which goes beyond commonly considered universality questions. Computational universality allows to implement NOT-type gates using pure-braids, and  also gives rise to injection weaves~\cite{Bonesteeletal05,Simonetal06},  but this is not necessarily sufficient to provide a distillation scheme along the lines explained here.   Another challenging
problem is the examination of the robustness of the scheme with respect to 
different initial states having composite topological charge. In practice, it may be necessary
to use a combination of the techniques presented here and more traditional methods relying on measurements and error-correction.

\vspace{0.1cm}


\begin{acknowledgements}
The author thanks John Preskill and Ben Reichardt for comments on an earlier draft. He also thanks the Erwin Schr\"odinger Institute and the Kavli Institute for Theoretical Physics for their hospitality. Support by  NSF grants PHY-0803371, PHY05-51164 and SNF grant PA00P2-126220 is gratefully acknowledged. 
\end{acknowledgements}

\end{document}